\documentclass[twocolumn,aip,apl,showpacs,amsfonts,amssymb,amsmath,superscriptaddress,groupedaddress]{revtex4}
\usepackage{graphicx}
\usepackage{amsmath}
\usepackage{amsfonts}
\usepackage{amssymb}
\usepackage{graphicx}

\begin{document}
\title{First principles study of electronic transport through a Cu(111)$|$graphene junction}
%\title{First principles study of electronic transport through a graphene$|$Cu(111) contact}
%\title{Electronic transport through Cu(111)-contacted graphene: \\ a finite bias {\em ab initio} study}
%\title{Transport through a graphene-metal interface: an efficient spin and momentum filter}

\author{Jesse Maassen}
\email{maassenj@physics.mcgill.ca}
\affiliation{Centre for the Physics of Materials and
Department of Physics, McGill University, Montreal, QC, Canada, H3A 2T8}

\author{Wei Ji}
\email{wji@ruc.edu.cn}
\affiliation{Department of Physics, Renmin University of China, Beijing 100872, China}
\affiliation{Centre for the Physics of Materials and
Department of Physics, McGill University, Montreal, QC, Canada, H3A 2T8}

\author{Hong Guo}
\affiliation{Centre for the Physics of Materials and
Department of Physics, McGill University, Montreal, QC, Canada, H3A 2T8}

%\author{Jesse Maassen}
%\email{maassenj@physics.mcgill.ca}
%\author{Wei Ji}
%\author{Hong Guo}
%\affiliation{Centre for the Physics of Materials and
%Department of Physics, McGill University, Montreal, QC, Canada, H3A 2T8}

\begin{abstract}
We report first principles investigations of the nonequilibrium
transport properties of a Cu(111)$|$graphene interface. The
Cu(111) electrode is found to induce a transmission minimum (TM)
located $-0.68\,{\rm eV}$ below the Fermi level, a feature
originating from the Cu-induced charge transfer resulting in {\em
n}-type doped graphene with the Dirac point coinciding with the TM. An applied
bias voltage shifts the {\em n}-graphene TM relative to the pure graphene TM 
and leads to a distinctive peak in the differential conductance indicating the
doping level, a characteristic not observed in pure graphene.
\end{abstract}

% 81.07.Bc Nanocrystalline materials
%73.50.-h Electronic transport phenomena in thin films
%31.15.Ac Ab initio group of atoms and clusters
%31.15.Ar Ab initio calculations
%81.07.Nb Molecular Nanostructures
%81.07.Lk Nanocontacts
%85.65.+h Molecular electronic devices
%72.10.Bg General formulation of transport theory
%72.20.Dp General theory, scattering mechanisms of conductivity
%73.23.-b Electronic transport in mesoscopic systems
%73.40.Sx Metal-semiconductor-metal structures
%73.63.-b Electronic transport in mesoscopic or nanoscale materials and structures

\maketitle

\begin{figure}
\includegraphics[width=7cm]{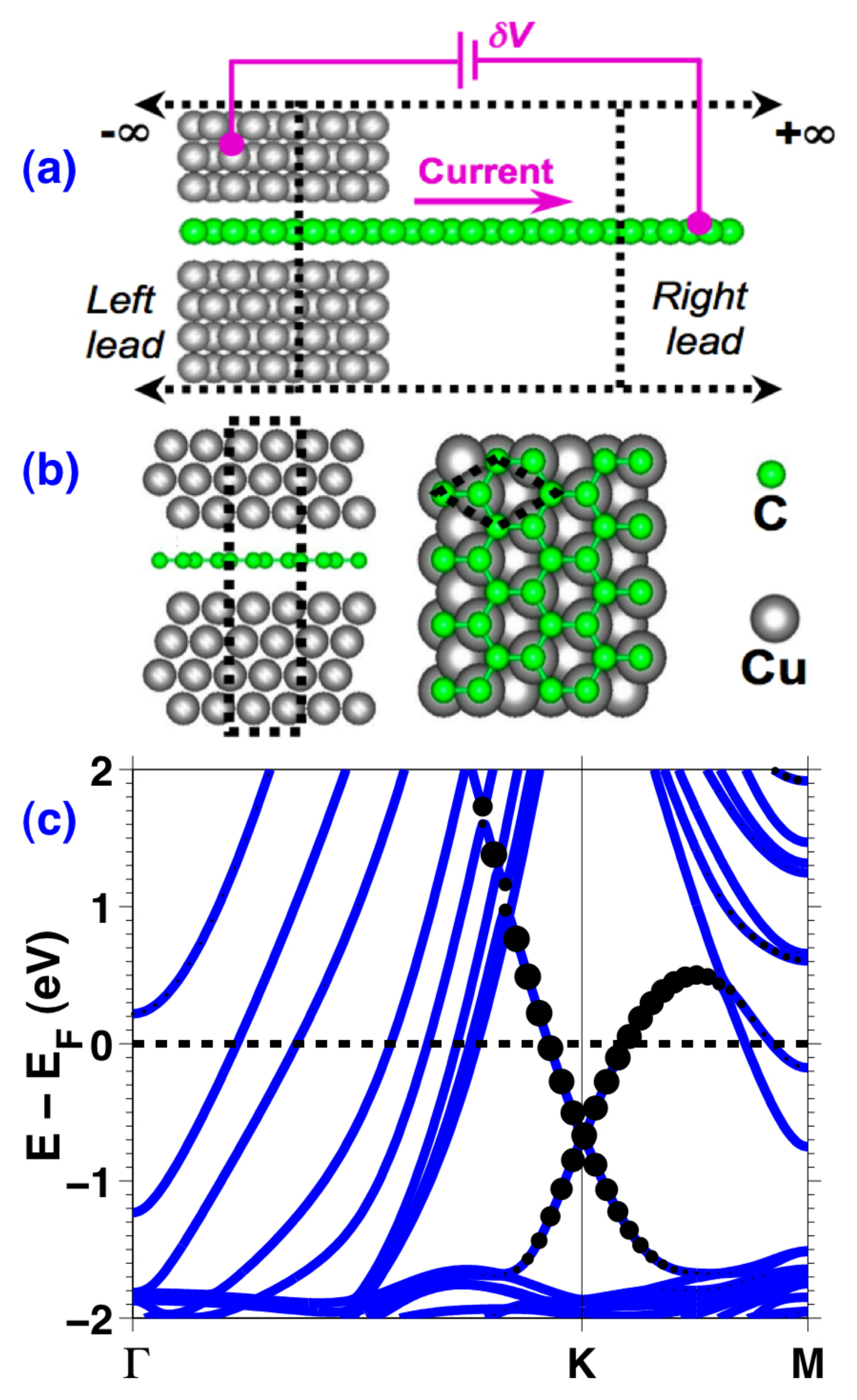}
\caption{(Color online) (a) Atomic structure of the Cu(111)$|$Gr interface.
The black dotted lines delimit the scattering region from the left lead (Cu-contacted
Gr) and right lead (pristine Gr) each extending to $\pm \infty$.
(b) Side- and top-view of the left lead. The optimized distance between the
Gr sheet and the Cu(111) surface is $\rm 2.95\, \AA$.
(c) Band structure of the left lead.
The states originating from the
hybrid Cu(111)$|$Gr system are plotted as solid blue lines, while the $\rm C(p_z)$ character
of the bands are superimposed as black circles.
The dotted black lines, shown in (b), indicate the supercell box used for
the electronic structure calculation. }
\label{fig1}
\end{figure}

Since the discovery of graphene (Gr), much research, fueled by the
advancement of scientific and technological progress at the
nanoscale, has focused on the exceptional intrinsic electronic
character of this two-dimensional material \cite{geim1}. Within the
context of nanoelectronics, utilizing Gr as an element in 
a device, e.g., electrode or channel, most likely requires interfacing it with
other materials, a process known to
potentially override/limit the unique properties of the Gr. In this
view, recent experimental \cite{exp} and theoretical
\cite{giovannetti1,theo,maassen,barraza1} studies have focused on the
contact properties of Gr-based devices and how the presence of
other materials can influence the global response of a device; a principle
used to demonstrate a high-speed Gr photodetector
\cite{ibm}. Thus in regard to transport, a physical understanding of
the Gr-metal contact, including the materials specific bonding, is
paramount.

With the advent of large-scale epitaxially grown Gr on Cu
\cite{fab}, relatively straightforward fabrication of macroscopic Gr
samples is possible, hence facilitating scientific and technological
purposes. This advancement also highlights the current importance of
Cu-Gr interaction. Whether etching part of the Cu substrate to realize
a device or transferring Gr to another substrate, some degree 
of Cu may remain in contact with the Gr. Thus a clear understanding of 
the influence of Cu on the electronic and transport 
characteristics of Gr is advantageous.

In this paper, we present a nonequilibrium {\em ab initio} study of
the transport properties of a Cu-Gr interface (as shown in Fig.\ref{fig1}(a),
hereafter referred to
as Cu$|$Gr). Our findings show that a Cu electrode induces a
transmission minimum (TM) located at $-0.68\,{\rm eV}$ below the
Fermi level ($\rm E_F$), denoted as TM1, in addition to the 
TM of pure Gr at $\rm E_F$ (TM2), both due to a vanishing 
density of states (DOS) at the Dirac
point. TM1 originates from a weak hybridization
between Gr and Cu, which (i) preserves Gr's linear dispersion bands
around the K point and (ii) induces a net charge transfer from Cu to Gr 
resulting in {\em n}-type doped Gr ({\em n}-Gr) \cite{giovannetti1}. 
Applying a bias shifts the position of the TMs relative to each other, 
leading to a peak in the differential conductance ($dI/dV$) 
that is otherwise not observed in pure Gr.

\begin{figure}
\includegraphics[width=7cm]{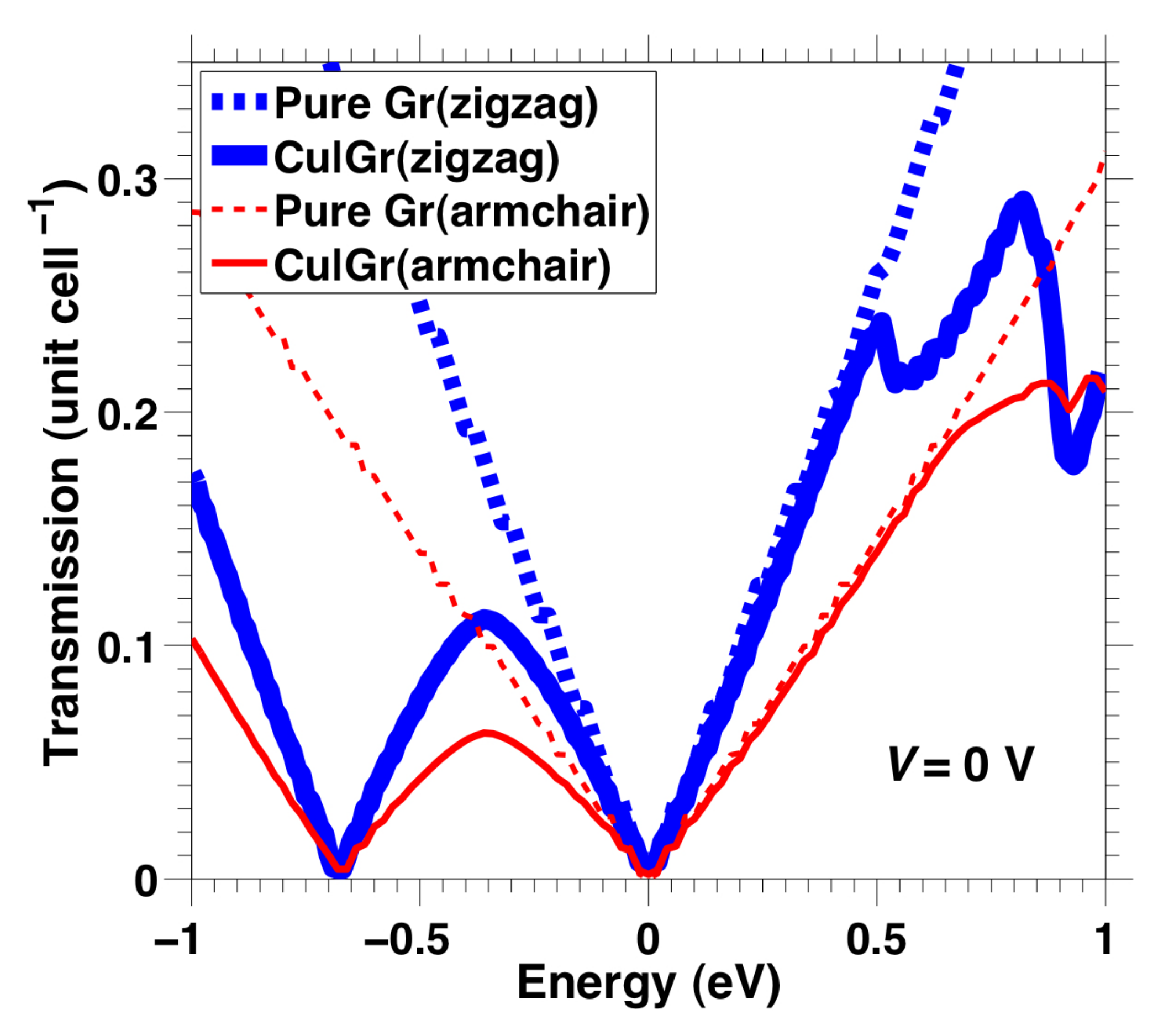}
\caption{(Color online) Transmission versus energy at equilibrium
($V=0\,{\rm V}$) as a function of Gr orientation. The solid thick
blue (thin red) line represents a Cu(111)$|$Gr interface where the
Gr is oriented such that transport occurs along the zigzag
(armchair) direction. The dashed thick blue (thin red) line shows
the transmission for pure zigzag-oriented (armchair-oriented) Gr.
The Fermi level is set to zero.} \label{fig2}
\end{figure}

The atomic coordinates of the Cu$|$Gr interface were optimized using
VASP \cite{vasp}, a density functional theory (DFT) package. The
nonequilibrium transport properties were calculated using 
{\sc MatDCal} \cite{matdcal}, which
combines nonequilibrium Green's functions (NEGF) with DFT to model
open systems in a two-probe geometry. For {\sc MatDCal}, we use an
optimized double-$\zeta$ polarized atomic basis set, the local
density approximation for exchange-correlation energy \cite{lda},
and norm-conserving non-local pseudopotentials \cite{pseudo} to
account for the nuclear and core potentials. Converged $k$-meshes,
including the important $\Gamma$, K and M points, 
were well tested and adopted in all 
calculations.

As illustrated in Fig.\ref{fig1}(a) the left and right leads, extending to
$\pm\infty$, are comprised of Cu-sandwiched Gr, simulating the
situation where a metallic contact is deposited on Gr, and pure Gr,
respectively. The system was chosen to
replicate the interface between a source/drain region of a Gr device
(Cu-coated Gr) and the pure Gr channel. Periodic boundary conditions
are imposed in the plane perpendicular to transport. 
Figure \ref{fig1}(b) shows the
atomic structure of the left lead of the Cu(111)$|$Gr interface. The
most stable configuration of Gr on Cu(111) corresponds to one
sub-lattice C atom situated directly above a Cu atom and the other C
atom located at the hollow site of the Cu(111) surface \cite{giovannetti1}. 
The supercell of the left lead for 
total energy and electronic structure calculations is
shown with the dotted box in Fig.\ref{fig1}(b), where the images of the Gr
sheets in neighboring supercells are separated by seven Cu(111)
layers.
By simultaneously relaxing the atomic coordinates to
forces $<0.01\,{\rm eV/\AA}$ and optimizing the supercell box size
by total energy minimization, the Gr-Cu(111) interlayer distance is
found to be $2.95\,{\rm \AA}$. 

The band structure of the left lead
is shown in Fig.\ref{fig1}(c). The blue lines represent the hybrid states 
of the Cu-Gr contact, while the black circles show the
$\rm p_z$ character of the C atoms corresponding to the Gr $\pi$-states.
From this projection, we observe that the Gr bands conserve a linear dispersion 
near the K point, as previously shown for Gr in contact with one Cu(111) 
surface \cite{giovannetti1}. 
The weak hybridization between Gr
and Cu merely shifts the conical Gr states down $-0.68\,{\rm eV}$
relative to $\rm E_F$ resulting in $n$-type doped Gr. This energy shift in
the Gr bands is somewhat larger than the value obtained for a single Cu(111)
surface with the same interlayer distance \cite{giovannetti1}.

\begin{figure}
\includegraphics[width=7.2cm]{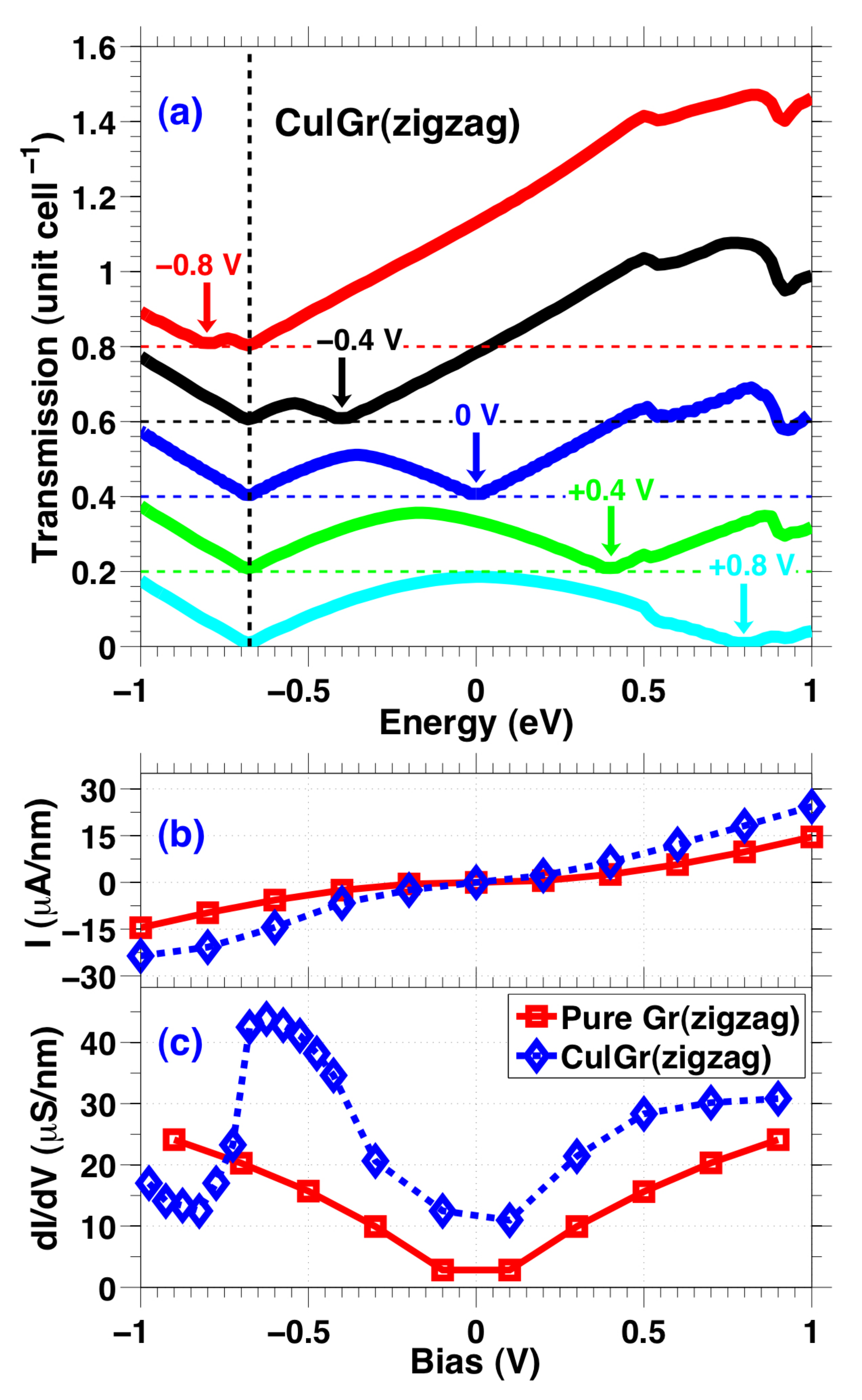}
\caption{(Color online) (a)
Transmission versus energy as a function of applied bias $V$ for
the Cu(111)$|$Gr(zigzag) interface. The transmission curves corresponding to
$V=+0.8\,{\rm V},\,+0.4\,{\rm V},\,0\,{\rm V},\,-0.4\,{\rm V},\,-0.8\,{\rm V},\,$
are offset by $0,\,0.2,\,0.4,\,0.6,\,0.8$ respectively, with
the zero plotted as a dashed horizontal line. The arrows
(black dashed vertical line) indicate the position of the
Dirac point of the pure Gr right lead ({\em n}-type Gr
left lead).
Current $I$ (b) and differential conductance $dI/dV$ (c) per unit
width versus $V$ for Cu(111)$|$Gr(zigzag) and pure Gr(zigzag).}
\label{fig3}
\end{figure}

Figure \ref{fig2} shows the
transmission coefficient ($T$) versus energy ($E$) as a function of
Gr orientation at equilibrium ($V=0$). A Gr sheet oriented such that
transport occurs along the zigzag or armchair direction is denoted
as Gr(zigzag) or Gr(armchair), respectively. The solid thick blue
(thin red) line corresponds to the Cu$|$Gr(zigzag)
(Cu$|$Gr(armchair)) interface. A TM (i.e., TM2) is observed at $\rm
E_F$ ($\rm E_F$ is shifted to zero), due to the vanishing density of
states (DOS) in the pure Gr right lead. Interestingly, a second
TM (i.e., TM1) is located at $-0.68\,{\rm eV}$. This feature also
originates from the near-zero DOS at the neutrality point of the
Cu-induced {\em n}-Gr in the left lead (see Fig.\ref{fig1}(c))
\cite{barraza1}. This indicates that the many available metallic
bands near the TMs, shown in the band structure, play a negligible
role in the electron flow of the junction. To
study the influence of the Cu electrode on the transport properties,
we calculate $T$ for pure Gr (dashed lines). We find that $T$ near
$\rm E_F$ is unaffected by the presence of the Cu electrode, while
the same is also true near the TM1. This confirms that
a Cu(111)$|$Gr interface can be effectively viewed as a {\em
n}-Gr$|$Gr junction. Finally, we consider the effect of Gr orientation
on $T$ and conclude that both curves are related by a constant
scaling factor. The ratio of $T_{\rm zigzag}/T_{\rm armchair}$ is
found to be equal to the ratio $W_{\rm armchair}/W_{\rm zigzag}$, where
$W$ is the width of the supercell used for transport.
Thus, $T$ (and hence current) per
unit width results in identical values for both zigzag and armchair
orientations.

In Fig.\ref{fig3}(a), we present $T$ versus $E$ as a function of
applied bias ($V$) for Cu$|$Gr(zigzag). A bias $V$ varies the
chemical potentials of the left and right leads as $\mu_L = {\rm
E_F}$ and $\mu_R={\rm E_F}+|e|V$, where $e$ is the electron charge. 
From Fig.\ref{fig3}(a), we see
that the TM1 is pinned at $-0.68\,{\rm eV}$ (shown with
the dashed vertical line), whereas the TM2 moves by an
amount $|e|V$ (indicated by the arrows). This is easily understood
because our bias is applied to the right Gr lead which shifts the
Dirac point there. By applying a bias one can shift the relative 
positions of the TM
features. From the $T$ calculated at finite $V$, we obtain the
current per unit width ($I$) for Cu$|$Gr(zigzag) and pure Gr(zigzag)
shown in Fig.\ref{fig3}(b). $I$ is calculated using the relation
\begin{equation}
I=\frac{2e}{hW} \int_{-\infty}^{+\infty} T(E,V)[f_L(E,\mu_R)-f_R(E,\mu_L)]\,dE \label{iv}
\end{equation}
where 
$h$ is Planck's constant, $W$ the unit cell
width and $f_{L,R}$ the Fermi-Dirac distribution functions. The
$I$-$V$ curves are smooth non-linear functions of $V$. To emphasize
the differences between the Cu$|$Gr and pure Gr, we calculate the
differential conductance per unit width $G\equiv dI/dV$, presented
in Fig.\ref{fig3}(c). Pure Gr results in a symmetric monotomic $G$,
whereas Cu$|$Gr shows a peak near $-0.63\,{\rm V}$ where both Dirac
points from the Gr and {\em n}-Gr coincide in energy. For $V$ greater than
$-0.63\,{\rm V}$, the TM1 plays no role in the values of $I$
and $G$, since it is out of range for the integration in Eq.(\ref{iv}) 
(from ${\rm E_F}+|e|V$ to $\rm E_F$, when $V<0$ and temperature is zero).
When $V<-0.63\,{\rm V}$, the TM1 enters the energy integration
range for $I$ and results in a rapid decrease in $G$. The maximum
derivative of $G$ provides the location of the conical point of the
{\em n}-Gr. This prediction should be experimentally detectable.

In summary, we performed first principles transport studies of the
Cu(111)$|$Gr interface. A Cu(111) electrode on Gr induces a second
conductance minimum near $-0.68\,{\rm eV}$ relative to $\rm E_F$, originating from the
vanishing density of states of the Cu-induced {\em n}-doped Gr. We
find a Cu$|$Gr interface can be effectively modelled as a {\em
n}-Gr$|$Gr junction due to a relatively weak interaction between Cu
and Gr. An applied bias shifts the positions of the conductance
minimums relative to each other, leading to a peak in the
differential conductance (indicating the doping level of the {\em
n}-Gr), a distinctive feature not observed in pure Gr. These results
provide a general picture of the influence of weakly interacting
metallic electrodes, including Al, Ag, Au and Pt \cite{khomyakov1}, on transport in Gr.

We acknowledge support from the FQRNT of Qu\'ebec, NSERC of Canada
and CIFAR. We thank RQCHP for computation facilities.

%\bibliography{tbref}

\begin{thebibliography}{00}
%
%%
\bibitem{geim1} A. K. Geim and K. S. Novoselov, Nature Mater. {\bf 6}, 183 (2007); A. H. Castro Neto, F. Guinea, N. M. R. Peres {\em et al.}, Rev. Mod. Phys. {\bf 81}, 109 (2009).
%
%
\bibitem{exp} E. J. H. Lee, K. Balasubramanian, R. T. Weitz {\em et al.}, Nature Nanotech. {\bf 3}, 486 (2008); T. Mueller, F. Xia, M. Freitag {\em et al.}, Phys. Rev. B {\bf 79}, 245430 (2009).
%
%
\bibitem{giovannetti1} G. Giovannetti, P. A. Khomyakov, G. Brocks {\em et al.}, Phys. Rev. Lett. {\bf 101}, 026803 (2008).
%
%
\bibitem{theo} Y. M. Blanter and I. Martin, Phys. Rev. B {\bf 76}, 155433 (2007); G. Liang, N. Neophytou, M. S. Lundstrom {\em et al.}, Nano Lett. {\bf 8}, 1819 (2008); R. Golizadeh-Mojarad and S. Datta, Phys. Rev. B {\bf 79}, 085410 (2009); Q. Ran, M. Gao, X. Guan {\em et al.}, Appl. Phys. Lett. {\bf 94}, 103511, (2009). 
%
%
\bibitem{maassen} J. Maassen, W. Ji and H. Guo, arXiv:--- (unpublished).
%
%
\bibitem{barraza1} S. Barraza-Lopez, M. Vanevic, M. Kindermann {\em et al.}, Phys. Rev. Lett. {\bf 104},
076807 (2010).
%
%
\bibitem{ibm} T. Mueller, F. Xia and P. Avouris, Nature Photonics {\bf 4}, 297 (2010).
%
%
\bibitem{fab} X. Li, W. Cai, J. An {\em et al.}, Science {\bf 324}, 1312 (2009).
%
%
\bibitem{vasp} G. Kresse and J. Furthmuller, Phys. Rev. B {\bf 54}, 11169 (1996); Comput. Mater. Sci. {\bf 6}, 15 (1996).
%
%
\bibitem{matdcal} J. Taylor, H. Guo and J. Wang, Phys. Rev. B {\bf 63}, 245407 (2001); {\bf 63}, 121104(R) (2001); D. Waldron, P. Haney, B. Larade {\em et al.}, Phys. Rev. Lett. {\bf 96}, 166804 (2006); J. Maassen, F. Zahid and H. Guo, Phys. Rev. B {\bf 80}, 125423 (2009).
%
\bibitem{lda} J. P. Perdew and A. Zunger, Phys. Rev. B {\bf 23}, 5048 (1981).
%
%
\bibitem{pseudo} N. Troullier and J. L. Martins, Phys. Rev. B {\bf 43}, 1993 (1991).
%
%
\bibitem{khomyakov1} P. A. Khomyakov, G. Giovannetti, P. C. Rusu {\em et al.}, Phys. Rev. B {\bf 79}, 195425 (2009).
%
\end{thebibliography}
%\bibliographystyle{}
%\nocite{*}

\end{document}